\documentclass[12pt]{article}
\usepackage{amsmath}
\usepackage{amssymb}

\newcommand{\BE}{\begin{equation}}
\newcommand{\EE}{\end{equation}}

\begin{document}

\setcounter{page}{0} \thispagestyle{empty}


\vspace*{1.5cm}

\begin{center}
{\large \bf Do $1/r$ potentials require massless particles ? }

\end{center}

\vspace*{1.5cm}

\renewcommand{\thefootnote}{\fnsymbol{footnote}}

\begin{center}
{\large M. Consoli}
 \\[0.3cm]
 {INFN - Sezione di Catania, I-95123 Catania, Italy}
\end{center}

\vspace*{0.5cm}

{

\vspace*{1.0cm}

\renewcommand{\abstractname}{\normalsize Abstract}
\begin{abstract}
Long-range $1/r$ potentials play a fundamental role in physics.
Their ultimate origin is usually traced back to the existence of
genuine massless particles as photons or gravitons related to
fundamental properties of continuum quantum field theories such as
gauge invariance. In this Letter, it is argued that, in principle,
an asymptotic, infinitesimally weak $1/r$ potential might also occur
in the cutoff version of a simple, one-component spontaneously
broken $\Phi^4$ theory, after taking into account the peculiar
nature of the zero-momentum limit of the connected scalar
propagator. Physical interpretation, phenomenological implications
and proposals for a new generation of lattice simulations are also
discussed.

\end{abstract}

\vspace*{0.5cm}
\begin{flushleft}
\end{flushleft}
\renewcommand{\thesection}{\normalsize{\arabic{section}.}}
\vfill\eject

{\bf 1.}~Long-range $1/r$ potentials play a very important role in
physics. Their ultimate origin is usually traced back to the
existence of genuine massless particles as photons or gravitons
related to fundamental properties of continuum quantum field
theories such as gauge invariance. In this Letter, it will be argued
that, in principle, an asymptotic, infinitesimally weak $1/r$
behaviour might also occur in the cutoff version of a simple,
one-component spontaneously broken $\Phi^4$ theory. To this end, one
has to take into account the peculiar nature of the zero 4-momentum
limit of the connected scalar propagator $G(p)$.

In fact, from the generally accepted "triviality" of the theory in
four space-time dimensions, one expects a gaussian structure of
Green's functions in the continuum limit. While this implies no
observable dynamics at any 4-momentum $p_\mu \neq 0$ and, on the
basis of Lorentz invariance, a free-field type form
$G^{-1}(p)=(p^2+m^2_h)$, one cannot exclude a discontinuity in the
zero-measure, Lorentz-invariant subset $p_\mu=0$. This plays a
fundamental role in translational invariant vacua characterized by
space-time constant expectation values of local operators such as
$\langle \Phi \rangle$.

For this reason, if there were arguments for the alternative
solution $G^{-1}(p=0)=0$, one might ask: could one consider a not
entirely trivial continuum limit where, still, $G^{-1}(p)=
(p^2+m^2_h)$ for any $p_\mu \neq 0$ but where there is a
discontinuity at $p_\mu=0$ and $G^{-1}(p=0)=0$ ? In this case, what
happens in the presence of an ultraviolet cutoff $\Lambda$ where one
expects instead a smooth behaviour ? At a certain point, for
sufficiently small ("infinitesimal") momenta, say $|p|\sim
m^2_h/\Lambda$, one should necessarily replace the standard massive
form $G^{-1}(p)\sim (p^2 + m^2_h) \to m^2_h$ with the different
alternative $G^{-1}(p) \to 0 $. Therefore, although the continuum
theory has only massive, free-field excitations, its cutoff version
would exhibit non-trivial qualitative differences, as weak
long-range forces, that cannot be considered uninteresting
perturbative corrections. This type of qualitative difference is the
main point of the present Letter.

In the following, I will first review the basic ingredients of the
problem. Some of these preliminary arguments are rather technical
and are listed as points {\bf 2}-{\bf 5} below. A reader who is only
interested in the main conclusions can simply look at the final
point {\bf 6}. Physical interpretation, possible phenomenological
implications and proposals for a new generation of lattice
simulations will also be discussed.

\vskip 10 pt

{\bf 2.}~Let us  start from lattice simulations. These were
performed \cite{further}, in the 4D Ising limit of the theory, to
objectively test the behaviour of the connected scalar propagator in
the broken phase. Differently from the symmetric phase at
$\langle\Phi\rangle=0$, where the simple massive picture works to
very high accuracy in the whole range of momenta, the results of the
low-temperature phase show unexpected deviations. Namely, when the
4-momentum $p_\mu\equiv ({\bf p},p_4) \to 0$, the propagator starts
to deviate from (the lattice version of) the form $1/(p^2+ {\rm
const.})$. By expressing the connected Euclidean propagator as \BE
\label{gg} G(p) = {{1}\over{p^2 +M^2(p^2)}} \EE these deviations can
be parameterized by using Stevenson's sensitive variable
\cite{stevensonnp} \BE \label{zeta}\zeta(p,m)\equiv (p^2 +m^2)G(p)
\EE In terms of this variable, the results can be summarized as
follows. One can first define a mass value $m\equiv m_h$ that well
describes the higher-momentum part of the propagator data, namely
those where one gets a remarkably flat $\zeta_{\rm latt}(p,m_h)
\lesssim 1$. In terms of this mass definition, the resulting
$\zeta_{\rm latt}(p,m_h)$ rapidly increase above unity in the $p_\mu
\to 0$ limit with a zero-momentum value \BE
 \zeta_{\rm latt}(0,m_h)=\left.{{{m^2_h}\over{M^2(0)}}}\right|_{\rm latt} \EE
that becomes larger and larger by approaching the continuum limit of
the lattice theory (compare Figs. 3, 4 and 5 of Ref.\cite{further}).
After the first indications of Ref.\cite{further}, Stevenson
\cite{stevensonnp} checked independently the existence of this
discrepancy by using different input masses for $\zeta(p,m)$ and
plotting the data in various ways. To this end, he used the lattice
data of Ref.\cite{balog} for the time slices of the connected
two-point correlator $C_1({\bf p}=0, t)\sim e^{-E(0)t}$ and
generated by Fourier transform equivalent data for the connected
scalar propagator $G({p})$. The resulting behaviour of $G(p)$ is in
complete agreement with the analogous plots obtained from
Ref.\cite{further} (compare Figs.6c, 7, 8 and 9 of
Ref.\cite{stevensonnp}).

The data also indicate that, by approaching the continuum limit, the
deviations from $\zeta_{\rm latt}(p,m_h) \lesssim 1$ become confined
to a smaller and smaller region of momenta near $p_\mu=0$. Thus, in
the continuum limit where the ultraviolet cutoff $\Lambda \to
\infty$, {\it both} $M(0)$ and the peculiar infrared region, say
$|p| \lesssim \delta$, where the propagator deviates from the simple
massive form, might vanish in units of $m_h$. In this scenario there
would be a hierarchy of scales $\delta \ll m_h \ll \Lambda$ such
that ${{\delta}\over{m_h}}\to 0$ when ${{m_h}\over{\Lambda}} \to 0$
(as for instance with the relation $\delta\sim m^2_h/\Lambda$).
Therefore, if $m_h$ were taken as the unit mass scale, the
deviations from a free-field massive behaviour would simply reduce
to the zero-measure set $p_\mu=0$. In this perspective, exact
Lorentz covariance would be recovered, since the value $p_\mu=0$
forms a Lorentz-invariant subset. Thus, the whole low-momentum
region would represent a typical example of "reentrant violation of
special relativity in the low-energy corner" \cite{volo}, namely one
of those peculiar infrared phenomena of cutoff theories.  In the
following I will now list three different theoretical arguments that
support an unconventional infrared behaviour and point to a similar
conclusion. \vskip 10 pt

{\bf 3.}~The first theoretical argument is based on the results of
Ref.\cite{dario}. There, one was studying the effective potential
$V_{\rm eff}(\phi)$ and the field strength $Z(\phi)$, as functions
of the background field $\phi$, at various values of the infrared
cutoff $k$. To this end, one starts from a bare action defined at
some ultraviolet cutoff $\Lambda$ and effectively integrates out
shells of quantum modes down to an infrared cutoff $k$. This
procedure generates a $k-$dependent effective action
$\Gamma_k[\Phi]$ that evolves into the full effective action
$\Gamma[\Phi]$ in the $k\to 0$ limit. In this approach, the relevant
quantities are the $k-$dependent effective potential $V_k(\phi)$ and
field strength $Z_k(\phi)$, which naturally appear in a derivative
expansion of $\Gamma_k[\Phi]$ around a space-time constant
configuration $\Phi(x)=\phi$.

By integrating numerically the coupled Renormalization-Group (RG)
equations for $V_k(\phi)$ and $Z_k(\phi)$, one finds the following
results. For not too small values of the infrared cutoff $k$, the
effective potential $V_k(\phi)$ remains a smooth, non-convex
function of $\phi$ as in the loop expansion. In this region of $k$
one also finds a field strength $Z_k(\phi)\sim 1$ for all values of
$\phi$.

However, a tiny momentum scale $\delta$ exists such that for $k
<\delta$ the effective potential $V_k(\phi)$ starts to flatten in an
inner region of $|\phi|$ while still matching with an outer,
asymptotic shape of the type expected in perturbation theory. The
flattening in the inner $|\phi|$-region, while reproducing the
expected convexity property of the exact effective potential, does
not correspond to a smooth behaviour. For such small values of $k$
there are large departures of $Z_k(\phi)$ from unity in the inner
$|\phi|-$region with a strong peaking at the end point
$|\hat{\phi}|=|\hat{\phi}(k)|$ of the flattening region. On the base
of the general convexification property, the $k \to 0$ limit of such
end point, $\hat{\phi}(0)$, coincides with one of the minima $\pm v$
of a suitable semiclassical, non-convex effective potential and is
usually taken as the physical realization of the broken phase.

Therefore, the fluctuations with $ |p| \leq \delta$ are
non-perturbative for values of the background field in the range
$-\hat{\phi}(|p|) \leq \phi \leq \hat{\phi}(|p|)$. In particular,
the very low-frequency modes with $|p| \to 0$ behave
non-perturbatively for all values of the background in the full
range $-v \leq \phi \leq v$ and thus cannot be represented as
standard weakly coupled massive states. Notice that the unexpected
effects show up with the emergence of the convexification process.
This is induced by the very long-wavelength modes that, so to speak,
"live" in the full region $-v \leq \phi \leq v$.

By itself, the existence of a non-perturbative infrared sector in a
region $0\leq |p| \leq \delta$ might not be in contradiction with
the assumed exact "triviality" property of the theory if, in the
continuum limit, the infrared scale $\delta$ vanishes in units of
the physical parameter $m_h$ associated with the massive part of the
spectrum. Again, this means to establish a hierarchy of scales
$\delta \ll m_h \ll \Lambda$ such that ${{\delta}\over{m_h}}\to 0$
when ${{m_h}\over{\Lambda}} \to 0$. Therefore, in units of $m_h$,
the region $0\leq |p| \leq \delta$ would just shrink to the
zero-measure set $p_\mu=0$ and one would be left with a massive,
free-field theory for all non-zero values of the 4-momentum.

{\bf 4.}~As a second theoretical argument, I will compare with
Stevenson's recent analysis \cite{recent} of the propagator in the
broken-symmetry phase. In his approach, a more faithful
representation of the true $\Phi^4$ interactions is obtained with
the non-local action \BE \int d^4x\int d^4y ~\Phi^2(x)
U(x-y)\Phi^2(y) \EE The kernel $U(x-y)$ contains, besides the
repulsive contact $\delta-$function term, say $U_{\rm core}(x-y)$,
an effective long-range attraction for $x \neq y$, say $U_{\rm
tail}(x-y)$. The latter, which is essential for a physical
description of spontaneous symmetry breaking as a true condensation
process \cite{mech}, originates from {\it ultraviolet-finite} parts
of one-loop Feynman graphs and has never been considered in the
perturbative RG$-$approach. Instead, by taking into account both
$U_{\rm core}$ and $U_{\rm tail}$ ( and avoiding double counting)
one can define a modified RG$-$expansion \cite{recent}, as in a
theory with two coupling constants. In the end, in the $\Lambda \to
\infty$ limit of the broken phase, the resulting connected Euclidean
propagator $G(p)$ approaches the standard free-field massive form
$G^{-1}(p)=(p^2+m^2_h)$ except for a discontinuity at $p_\mu=0$
where $G^{-1}(p=0)=0$. This type of structure, implying the
existence of a branch of the spectrum whose energy $E({\bf p})\to 0$
in the ${\bf p}\to 0$ limit, would indeed support the previous idea
that, at least for the continuum theory, all deviations from the
massive behaviour are at $p_\mu=0$. \vskip 10 pt

{\bf 5.}~ Finally, as a third theoretical argument, I emphasize that
the possibility $G^{-1}(p=0)=0$ is also in agreement with the
analogous indication of Ref.\cite{consoli} that, in the
broken-symmetry phase, $G^{-1}(p=0)$ is a {\it two-valued} function
that, in addition to the standard value $G^{-1}_a(p=0)=m^2_h$,
includes the solution $G^{-1}_b(p=0)=0$ as in a massless theory. To
this end, it becomes crucial to take the $\phi \to \pm v$ limit of
the broken phase by first including the one-particle reducible
tadpole graphs where zero-momentum propagator lines are attached to
the one-point function $\Gamma_1(p=0)=V'_{\rm NC}(\phi)$, the first
derivative of the standard non-convex effective potential $V_{\rm
NC}(\phi)$ of the loop expansion. By implicitly assuming the
regularity of the zero-momentum propagator, these graphs are usually
ignored {\it at} $\phi=\pm v$ where $V'_{\rm NC}(\pm v)=0$. Thus,
$G^{-1}(p)$ is identified with the 1PI two-point function
$\Gamma_2(p)$, whose zero-momentum value $\Gamma_2(p=0)$ is nothing
but $V''_{\rm NC}(\pm v)$, a positive-definite quantity. On the
other hand, by allowing for a singular $G(p=0)$, one is faced with a
completely different diagrammatic expansion and thus the simple
picture of the broken phase as a pure massive theory, based on the
chain \BE G^{-1}(p=0)=\Gamma_2(p=0)=V''_{\rm NC}(\pm v)
> 0\EE breaks down.

It is interesting that, as in point {\bf 3} above, one can find a
relation with the convexity property of the exact effective
potential. In fact, the existence of the two solutions for
$G^{-1}(p=0)$ at $\phi=\pm v$ can also be derived by evaluating in
the saddle point approximation the generating functional $W[J]$ for
a constant source and taking the double limit where $J \to \pm 0$
and the space-time volume $\Omega \to \infty$ \cite{consoli}.

As such, the two solutions admit a geometrical interpretation in
terms of left and right second derivatives of the exact, Legendre
transformed effective potential $ V_{\rm LT}(\phi)$ . This is convex
downward  and is not an infinitely differentiable function when
$\Omega \to \infty$ \cite{syma}. These non-trivial differences
should induce to check those physical aspects of the spontaneously
broken phase, such as the mass spectrum, that depend crucially on
the identification $V_{\rm eff}(\phi) \equiv V_{\rm NC}(\phi)$. In
particular, the $k-$dependent effective potential $V_k(\phi)$,
obtained by integrating out shells of quantum modes down to some
infrared cutoff $k$ and mentioned in Sect.{\bf 3}, is clearly
approaching convexity in the $k \to 0$ limit. Therefore, this well
defined theoretical construction supports the identification of $
V_{\rm LT}(\phi)$ as the true effective potential in the
infinite-volume limit of the theory.

\vskip 10 pt

{\bf 6.}~It is conceivable that the subtleties of $G(p)$ at
$p_\mu=0$ might have been missed in most conventional approximation
schemes. At the same time, the possibility of an infrared sector
which is richer than expected has far reaching phenomenological
implications. To see this, let us first summarize the results of
sects. {\bf 2}-{\bf 5} as follows : by assuming a continuum limit
where Lorentz invariance and "triviality" hold exactly, a possible
deviation from the entirely trivial, massive free-field limit, with
$G^{-1}(p)=(p^2+m^2_h)$ identically, can only have the form of a
discontinuity at  $p_\mu= 0$ where $G^{-1}(p)=0$.

Starting from this observation, in the presence of a finite
ultraviolet cutoff $\Lambda$, where one expects instead a smooth
behaviour, one can try to construct a not entirely trivial $G(p)$ as
a smooth interpolation between these two distinct propagator forms
of the continuum theory, say \BE \label{inter}G^{-1}(p)=(p^2 +m^2_h)
f(p^2/\delta^2) \EE The function $f$ refers to some infrared
momentum scale $\delta\neq 0$ (with $\delta/m_h\to 0$ when
$m_h/\Lambda \to 0$) in such a way that \BE \lim_{\delta\to
0}f(p^2/\delta^2)=1~~~~~~~~~~~~~(p_\mu \neq 0)\EE with the only
exception \BE \lim_{p_\mu\to 0}f(p^2/\delta^2)=0~~~~~~~~~~~~~\EE
(think for instance of $f(x)=\tanh(x)$, $f(x)=1 - \exp(-x)$,
$f(x)=x/(1+x)$,...). In the following I will adopt Eq.(\ref{inter}).
However, as one can easily check, there would be no significative
change by employing the alternative form $G^{-1}(p)=p^2 +m^2_h
f(p^2/\delta^2)$. In fact, analogous results would persist in any
cutoff version where the function $M^2(p^2)$ of Eq.(\ref{gg})
vanishes for $p_\mu \to 0$. Notice that, by adopting
Eq.(\ref{inter}), one simply finds
$f(p^2/\delta^2)=\zeta^{-1}(p,m_h)$ in terms of Stevenson's
$\zeta$-function (\ref{zeta}).

To understand what kind of instantaneous potential $V({\bf r})$ in
coordinate space is associated with such propagator for the scalar
field, one has to consider the standard Fourier transform of the
zero-energy propagator $G({\bf p},p_4=0)$ \BE \label{fourier} D({\bf
r})= \int {{d^3p}\over{(2\pi)^3}}
 {{e^{i {\bf p}\cdot{\bf r}}}\over{ ({\bf p}^2 + m^2_h) f({\bf p}^2/\delta^2) }}\EE
that in the case of the one-photon exchange, $G({\bf
p},p_4=0)=1/{\bf p}^2$, gives a $1/r$ potential.

Now, a straightforward replacement $ f({\bf p}^2/\delta^2) =1$ would
produce the well known Yukawa potential $e^{-m_h r}/r$. However, if
we consider the finite-cutoff theory, we have to take into account
the region $ {\bf p}^2 \ll \delta^2$ where the relevant limiting
relation is rather \BE \lim_{{\bf p} \to 0}f({\bf
p}^2/\delta^2)=0~~~~~~~~~~~~~\EE For this reason, since the dominant
contribution for $r \to \infty$ comes from ${\bf p}=0$, where the
denominator in (\ref{fourier}) vanishes, there would be long-range
forces that have never been considered. In this case, by expanding
around ${\bf p}=0$ and replacing \BE \label {simple} f({\bf
p}^2/\delta^2) \sim {{{\bf p}^2}\over{ \delta^2}}f'(0) \EE one
obtains the two leading behaviours \BE \label{fourier2} \lim_{ {\bf
p} \to 0}~ \left( {\bf p}^2 + m^2_h\right) f({\bf p}^2/\delta^2)
\sim {\bf p}^2 {{m^2_h}\over{\delta^2 }} f'(0) \EE and \BE
\label{fourier3} \lim_{ {\bf p}^2 \to \infty}~ \left({\bf p}^2 +
m^2_h\right) f({\bf p}^2/\delta^2) \sim {\bf p}^2 \EE Therefore, on
the basis of the Riemann-Lebesgue theorem on Fourier transforms
\cite{goldberg} (see the Appendix), whatever the detailed form of
$f(x)$ at intermediate $x$,  the leading contribution at
asymptotically large $r$ will be 1/r. One thus gets \BE
\label{asy}\lim_{r \to \infty}~D({\bf r})= D_{\infty}({\bf r})={{
\delta^2}\over{f'(0)m^2_h}}~{{1}\over{4\pi r}}\EE all dependence on
the interpolating function being contained in the factor
$f'(0)={\cal O}(1)$.

To put some numbers (in units $\hbar=c=1$), let us consider for
definiteness the scenario $\delta\sim m^2_h/\Lambda$
\cite{consoli2}. This is motivated by a description of spontaneous
symmetry breaking as a true condensation process and by the
identification of $\delta$ as the momentum scale below which
collective oscillations of the condensate starts to propagate
\cite{note1}. Thus, if $m_h$ were around the Fermi scale and
$\Lambda$ around the Planck scale, $\delta$ would be around
$10^{-5}$ eV. For this particular case, let us compute the
asymptotic potential between two fermions $i$ and $j$ of masses
$m_i$ and $m_j$ that in the Standard Model couple to the singlet
Higgs boson with strength $y_i=m_i/v$ and $y_j=m_j/v$. Besides the
short-distance Yukawa potential governed by the Fermi constant
$G_F\equiv 1/v^2$ \BE \label{fourier5}V_{\rm yukawa}({\bf r}) =
-{{G_F m_i m_j}\over{4\pi r}}e^{-m_h r}\EE (that dominates for $r
\lesssim 1/m_h$) they would feel  the asymptotic potential
associated with Eq.(\ref{asy}). This can be conveniently expressed
as \BE \label{fourier6} \lim_{r \to \infty}~V({\bf
r})=V_{\infty}({\bf r}) = -{{G_{\infty} m_i m_j}\over{4\pi r}} \EE
with the effective coupling  \BE G_{\infty}=
{{\delta^2}\over{f'(0)m^2_h }}~G_F \sim 10^{-33}G_F \EE Strictly
speaking, this asymptotic potential represents a cutoff artifact
since the continuum theory has only massive, free-field excitations,
with the only exception of a discontinuity at $p_\mu=0$ where
$G^{-1}(p)=0$. At least, this seems the only possible remnant of
symmetry breaking allowed by exact Lorentz invariance and
"triviality". However in the cutoff theory, where one expects a
smooth behaviour, the deviation from the massive form will
necessarily extend, from the zero-measure set $p_\mu=0$, to a tiny
momentum region $\delta$. It is this momentum region, that vanishes
in the continuum theory but remains finite in the cutoff theory, to
produce the long-range $1/r$ potential of strength $\delta^2/m^2_h$.
Since in this momentum region the propagator looks like in a
massless theory, the answer to the question posed in the title of
this Letter depends on the personal taste, even though there are no
genuine massless particles (i.e. with propagator $1/p^2$ in the
whole range of momenta). Notice that, in the context of a condensate
physical picture, the idea of long-range $1/r$ interactions in
$\Phi^4$ theory below the condensation temperature was also
considered in Ref.\cite{ferrer} by following a different approach.

A possible physical interpretation of the phenomenon is the
following. By representing the broken-symmetry phase as a physical
condensate, one is naturally lead to consider superfluid $^4$He, the
physical system that is usually considered as a non-relativistic
realization of $\Phi^4$. One can thus try to understand the
double-valued nature of the zero-momentum connected propagator in
the broken phase in analogy with Landau's original idea
\cite{landau} of two different branches in the energy spectrum of
$^4$He, namely gapless density oscillations (phonons) and massive
vortical excitations (rotons) \cite{note2}. Experiments however have
shown that these two different branches actually merge into a single
energy spectrum, a sort of "hybrid" that smoothly interpolates
between the two different functional forms. In our case, the
interpolating propagator produces similar effects.

One may object that Eq.(\ref{simple}) might be too simple.  In
principle, the function $f(p^2/\delta^2)$, for $p_\mu \to 0$, might
vanish as $(p^2/\delta^2)^{1+\eta}$, where $\eta$ plays the role of
an anomalous dimension and might be needed for a proper matching of
the inverse propagator in the infrared region. In this case, the
asymptotic $1/r$ potential in coordinate space would exhibit
corrections proportional to $(r\delta)^\eta$.

Another possible objection is that a scale $\delta\sim 10^{-5}$ eV
is probably ruled out by experiments. Thus, in the scenario
$\delta\sim m^2_h/\Lambda$, to get a sufficiently small strength,
one should take a $\Lambda$ which is larger than the Planck scale or
a $m_h$ well below 300 GeV or both. However, comparison with
experiments represents a separate issue. If some assumption behind
the above numerical analysis is in conflict with phenomenology,
still, the basic ambiguity of $G(p)$ at $p_\mu =0$ remains a
peculiarity of the broken-symmetry vacuum and represents a challenge
for any consistent cutoff version of the theory.

In conclusion, for its conceptual relevance and the potential
phenomenological implications, it seems worth to further sharpen our
understanding of the low-momentum region of spontaneously broken
$\Phi^4$ theories. In particular, with a new generation of lattice
simulations one should study the $p_\mu \to 0$ limit of the
connected propagator on much larger lattices and try to determine
the interpolating function $f(p^2/\delta^2)$ in Eq.(\ref{inter}). By
reaching the critical region $ |p| \lesssim \delta$, the deviations
from the pure massive behaviour $f \sim 1$ (that remain below 30$\%$
on the lattices used so far \cite{further,stevensonnp}) should
become macroscopical. In the scenario $\delta \sim m^2_h/\Lambda$,
by using the relation \cite{brahm} $\Lambda \sim
{{4.893(3)}\over{a}}$ to relate the ultraviolet cutoff of a
$\Phi^4_4$ theory to the lattice spacing $a$, and setting
$\delta={{2\pi}\over{L_{\rm min}}}$, this means a minimal lattice
size \BE {{ L_{\rm min}}\over{a}} \sim {{30.74}\over{(m_ha)^2}} \EE
For mass values in the scaling region, one gets ${{L_{\rm
min}}\over{a}} \sim $ 123, 192, 342, 769, 3074 for $m_h a$=0.5, 0.4,
0.3, 0.2 , 0.1 respectively. Four-dimensional lattices with ${{
L}\over{a}}={\cal O}(100)$ should be attainable with the present
supercomputers.

\vskip 5 pt

\centerline{{\bf Acknowledgments}} \vskip 5 pt I thank M. Baldo, P.
Cea, L. Cosmai, P. M. Stevenson and D. Zappal\`a for useful
discussions and collaboration.

\begin{center}
{\large{\bf Appendix}}

\end{center}
Let us consider a function $F(q^2)$ that exhibits the two asymptotic
trends ($\alpha >0$, $A>0$)\BE \lim_{q \to 0}{{F(q^2)}\over{q^2}} =
\alpha + {\cal O} (q^2)~~~~~~~~~~~~~~~~~~~~\lim_{q \to
\infty}{{F(q^2)}\over{q^2}} = A + {\cal O} (1/q^2) \EE so that \BE
\int^{\infty}_{0}{{q~ dq}\over{|F(q^2)|}}= + \infty \EE I will
assume the general requirements needed for the existence of the
Fourier transform \BE I(r)=\int {{d^3q}\over{(2\pi)^3}} {{e^{i {\bf
q}\cdot{\bf r}} }\over{F(q^2)}}= {{1}\over{2\pi^2 r}}
\int^{\infty}_{0} {{q~ dq}\over{F(q^2)}}~ \sin(qr) \EE Our aim is to
determine the leading behaviour of $I(r)$ for $r \to \infty$. To
this end, one can introduce a momentum scale $\delta$ and decompose
$I(r)$ as \BE I(r)=I_1(r)+I_2(r)+I_3(r) \EE where \BE I_1(r)=
{{1}\over{2\pi^2
r}}\int^{\infty}_{0}{{1}\over{q}}\left({{q^2}\over{F(q^2)}} - {{q^2
+\delta ^2 }\over{Aq^2 + \alpha \delta^2}}\right)\sin (qr)~dq \EE\BE
I_2(r)= {{1}\over{2\pi^2 r}}\int^{\infty}_{0}
{{1}\over{q}}\left({{q^2 +\delta ^2 }\over{Aq^2 + \alpha \delta^2}}-
{{1}\over{\alpha}}\right)
 \sin (qr)~dq
\EE
 \BE I_3(r)=
{{1}\over{2\pi^2\alpha r}}\int^{\infty}_{0}{{\sin (qr)}\over{q}}~dq=
{{1}\over{4\pi\alpha r}} \EE By introducing the function \BE g(q)
={{1}\over{q}}\left({{q^2}\over{F(q^2)}}- {{q^2 +\delta ^2
}\over{Aq^2 + \alpha \delta^2}}\right) \EE one gets \BE
\int^{\infty}_{0} dq~ |g(q)| < + \infty \EE (i.e. $g(q)~\epsilon~
L^{(1)} $). For this reason by defining \BE \hat{g}(r)=
\int^{\infty}_{0}g(q) \sin (qr)~dq \EE one finds \BE \lim_{r \to
\infty} \hat{g}(r) =0 \EE for the Riemann-Lebesgue theorem
\cite{goldberg}. Thus $I_1(r)=\hat{g}(r)/(2\pi^2r)$ vanishes faster
than $1/r$ when $r\to \infty$. On the other hand, one also finds \BE
I_2(r)={{1}\over{2\pi^2 r}}{{\alpha-
A}\over{\alpha}}\int^{\infty}_{0} {{q}\over{Aq^2 + \alpha \delta^2}}
\sin (qr)~dq={{\alpha-
A}\over{A\alpha}}{{e^{-\sqrt{{{\alpha}\over{A}}}~\delta
r}}\over{4\pi r}} \EE Therefore, in the $r \to \infty$ limit, the
leading behaviour of $I(r)$ is given by $I_3(r)$.


\begin{thebibliography}{10}
\bibitem{note1}
In the picture of Ref.\cite{mech} the scalar condensate is
characterized by two basic quantities, the number density $n$ of the
elementary condensed spinless quanta and their scattering length
$a$. One can then show that these two quantities are connected to
the mass parameter $m_h$ through the relation $m_h\sim \sqrt{na}$.
Thus, the `triviality' limit, where $a m_h \to 0$, can be simulated
by an ultraviolet cutoff $\Lambda \sim 1/a$ such that
$m^2_h/\Lambda^2 \sim na^3 \to 0$. Furthermore, by noticing that in
all known media collective density oscillations propagate over
distances larger than the mean free path $r_{\rm mfp}\sim
1/(na^2)\sim \Lambda/m^2_h$ for the elementary constituents, one
obtains  the order of magnitude relation $\delta \sim 1/r_{\rm
mfp}\sim m^2_h/\Lambda$.
\bibitem{note2}
This interpretation is also supported by the results of
Ref.\cite{hydro}. There, by using quantum hydrodynamics, the mass
parameter $m_h\sim \sqrt{na}$ was shown to be proportional to the
energy gap for vortex formation in a superfluid medium possessing
the same constituents and the same density as in the condensate
picture of $\Phi^4$.
\end{thebibliography}
\end{document}